\documentclass[]{aa}

\usepackage[varg]{txfonts}   
\usepackage{graphicx}        
\usepackage{amsmath}         
\usepackage{xcolor}          
\usepackage{hyperref}        
\usepackage{epstopdf}
\usepackage{fancyhdr}

\hypersetup{
    colorlinks=true,          
    linkcolor=blue,          
    citecolor=blue,          
    urlcolor=blue            
}

\newcommand{\angstrom}{\text{\normalfont\AA}}
\newcommand\gaia{\textit{Gaia }}

\newcommand\RGaia{$11\,500$}
\newcommand\RGALAH{$28\,000$}

\newcommand\Nvbroad{$3\,524\,677$}
\newcommand\NGALAHvsini{$581\,149$}

\newcommand\NvbroadTwoStripes{$81\,371$}

\begin{document}

\title{Gaia \texttt{vbroad}---the Spectral-Line Broadening, and Binarity}
\titlerunning{\gaia \texttt{vbroad} and binarity}

\author{
    E. Hadad$^1$, T. Mazeh$^1$, S. Faigler$^1$
    and
    A.G.A. Brown$^2$
}

\institute{
    School of Physics and Astronomy, Tel Aviv University, Tel Aviv, 6997801, Israel\\
    \email{eyalhadad@mail.tau.ac.il}
    \and
    Leiden Observatory, Leiden University, Einsteinweg 55, 2333 CC
Leiden, The Netherlands \\ 
    \email{brown@strw.leidenuniv.nl}
}

\abstract{
The \gaia DR3 catalog includes line-broadening measurements (\texttt{vbroad}) for \Nvbroad~stars. We Concentrate here on the low-mass main-sequence (MS) sub-sample of the catalog, with $(G_\mathrm{BP}-G_\mathrm{RP})_0$ in the range of 1--1.6, which includes \NvbroadTwoStripes~sources. The colour-magnitude diagram of the sample displays two distinct strips, the brighter of which is probably mostly composed of unresolved binaries, with mass ratios close to unity. We show that the suspected binary sub-sample displays a larger \texttt{vbroad} distribution, which we attribute to the unresolved absorption lines of the two components of each binary. A similar effect is seen in the GALAH data.

}


\maketitle
\section{Introduction}
The \gaia space mission \citep{Gaia16} carries
a spectrometer with intermediate resolving power, R $\approx$ \RGaia, covering
the $846$--$870$ nm range, with the primary goal of measuring the radial velocity (RV) of the bright sources 
\citep{cropper18},\footnote{https://gea.esac.esa.int/archive/documentation/GDR3/index.html} 
 down to magnitude $G_\mathrm{RVS}$ $\sim16$ \citep[]{katz23}.
The spectroscopic pipeline \citep{Sartoretti18}
also derives a line-broadening
parameter that estimates the broadening of the absorption lines relative to the pertinent template, yielding for \gaia DR3 \citep{GaiaDR323}
\texttt{vbroad} values of \Nvbroad~sources \citep[]{Fremat23}.\footnote{https://gea.esac.esa.int/archive/}

Despite the relatively low resolution of the 
spectrograph, the \texttt{vbroad} values can be used to estimate the stellar rotational broadening \citep[e.g.,][]{gilhool19}, as discussed by \citet[]{Fremat23}, which in turn can help estimate the stellar rotational period, in case we know the stellar radius \citep[e.g.,][]{kiman24} and the rotational axis inclination. Stellar rotation was derived in the last years from the observed photometric modulation \citep[e.g.,][]{mcquillan14, StellarRotationOgle24, StellarRotationTess24}, utilizing large photometric surveys, carrying astrophysical information on the different types of stars \citep[e.g.,][]{Buder21, Xiang22}. The \texttt{vbroad} database is another independent avenue to obtain stellar rotation periods, albeit less accurate. 

Another large sample of stellar broadening is included in the GALAH DR3 database \citep[][]{Buder21}, based on observations of the HERMES spectrometer amounted on the Anglo-Australian Telescope (ATT). The \texttt{vbroad}, given for \NGALAHvsini~sources,\footnote{https://vizier.cds.unistra.fr/viz-bin/VizieR-3?-source=J/MNRAS/506/150/stars} is one of its stellar parameters. With HERMES, the AAT achieves a median spectral resolution of \RGALAH~to create a sample surpassing \gaia's quality. HERMES covers a non-contiguous range of $1\,000\angstrom$, spanning across the blue ($4\,713$--$4\,903\angstrom$), green ($5\,648$--$5\,873\angstrom$), red ($6\,478$--$6\,737\angstrom$), and infrared ($7\,585$--$7\,887\angstrom$) wavelengths. \texttt{vbroad} is given for sources up to $V\sim20$, $92\%$ of which are brighter than $V=14$ \citep{Barden2010, Heijmans2012, Farrell2014, Sheinis2015}. 

In this paper, we study the impact of the binarity of the main-sequence (MS) \gaia sources on the derived line width---\texttt{vbroad}. We identified the binaries by their brightness excess relative to their colour on the \gaia colour-magnitude diagram (CMD) \citep[e.g.,][]{donada23, wallace24}. 
\cite{Kovalev2024, Kovalev2022} used a somewhat similar approach, identifing \textasciitilde$10^4$ double-lined spectroscopic binary (SB2) candidates in LAMOST spectra by their larger rotational velocity values.

Section~\ref{sec:Identifying 2 Stripes} shows that the sample of late MS \gaia sources with derived \texttt{vbroad} values displays two strips. The upper strip is mostly composed of unresolved binaries with relatively luminous secondaries, displaying a significantly larger \texttt{vbroad} distribution. 
Section~\ref{sec:GALAH} repeats the analysis for a sample of stars with GALAH \texttt{vbroad}, showing similar results. Section \ref{sec:discussion} discusses and concludes our findings. 

\section{Identifying two CMD Strips in the late-type MS 
{\it Gaia} stars with \texttt{vbroad} measurements}
\label{sec:Identifying 2 Stripes}

\begin{figure}
	\includegraphics[width=\columnwidth]{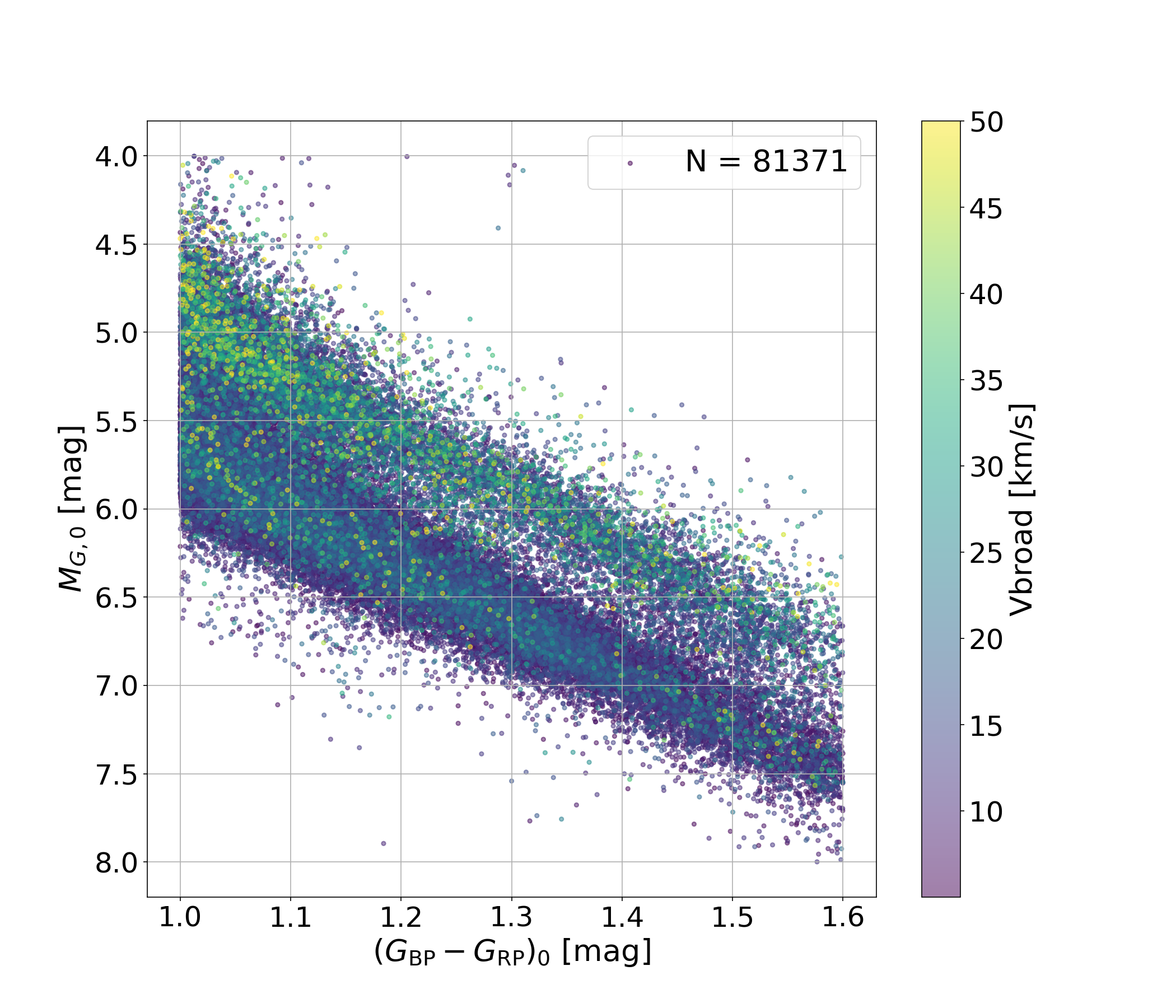}
\caption{Our \texttt{vbroad} sample, for sources with given extinction coefficients, $\varpi > 0$, $\sigma_{\varpi}/\varpi < 10\%$, $(G_\mathrm{BP}-G_\mathrm{RP})_0$ $\in [1, 1.6]$ and $M_{G,0}$ $\in [4, 8]$, colour-coded for \texttt{vbroad}. The $215$ sources with \texttt{vbroad} well above $50$ km/s are rounded to that value in the colour coding; of these, $204$ are located on the upper strip.}
    \label{fig:2 stripes color coded}
\end{figure}

Out of the  original \texttt{vbroad} sample, which consists of \Nvbroad~sources, we  consider only sources with 
$\varpi > 0$ and $\sigma_{\varpi}/\varpi < 10\%$,  
leaving $3\,238\,383$ sources, from which we only consider those with given extinction coefficients, reducing our population size to $1\,834\,196$. \\
To focus on the late MS stars of the sample, we consider a restricted sample of stars within an extinction-corrected colour of $(G_\mathrm{BP}-G_\mathrm{RP})_0$ range of $1$--$1.6$, and extinction-corrected absolute $G$ magnitude ($M_{G,0}$) of $4$--$8$, leaving us with \NvbroadTwoStripes~stars, plotted in  
Fig.~\ref{fig:2 stripes color coded}, colour-coded by the \texttt{vbroad} values. The diagram clearly displays two parallel strips, where the upper strip displays larger values of \texttt{vbroad}. This strip is probably composed of unresolved binaries \citep{freund24,
phillips24, wallace24}, which are more luminous than their primaries, due to the brightness contribution of their secondaries.
.

\begin{figure}	
\includegraphics[width=\columnwidth]{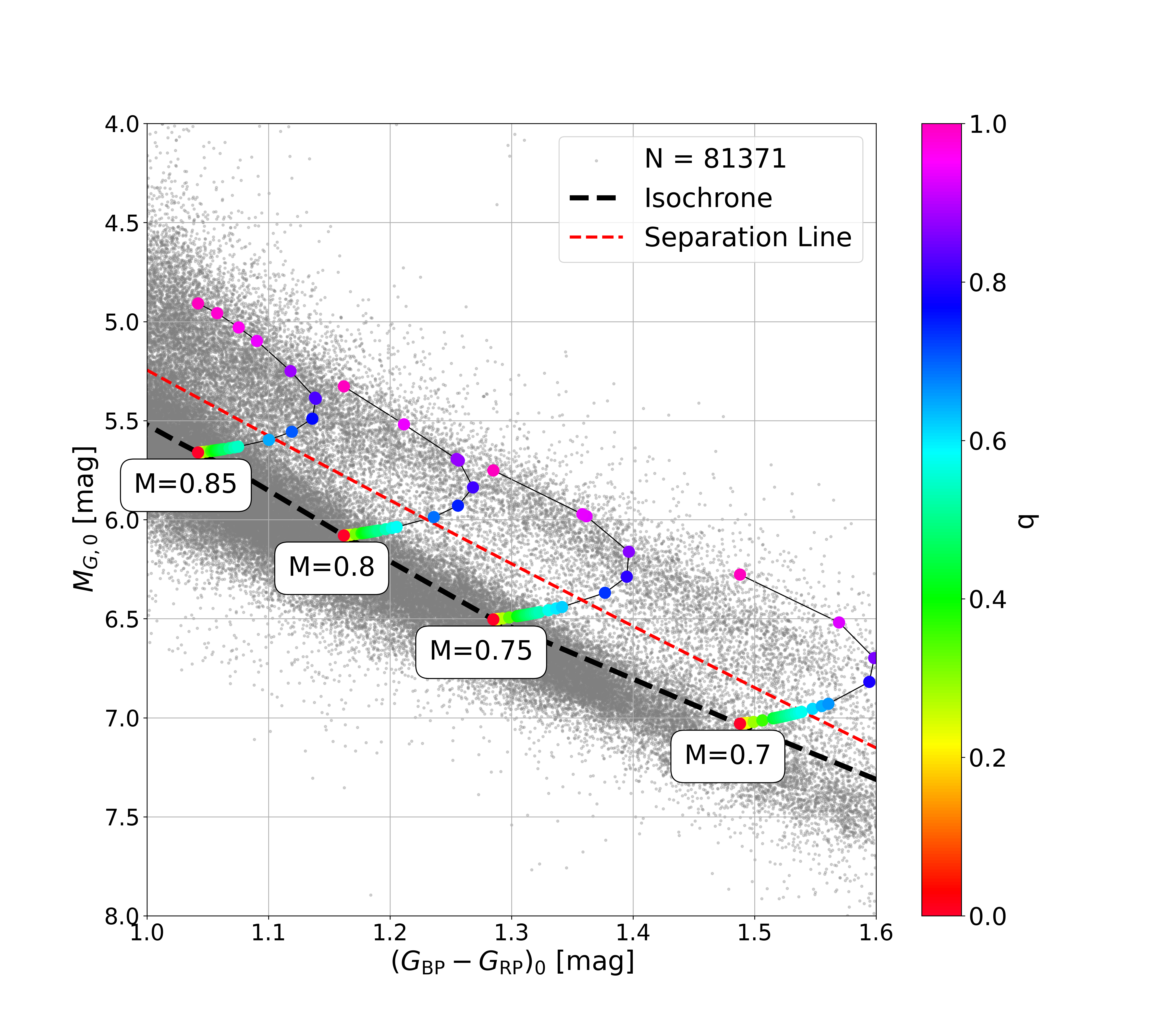}
\caption{
Same as Fig~\ref{fig:2 stripes color coded}, over-plotted with isochrones with an age of 4.5 Gyr and metallicity [Fe/H] = 0. Four primary masses (0.7, 0.75, 0.8 \& 0.85 solar mass) are positioned on the isochrone. For each isochrone, binaries of different mass ratios are simulated and colour-coded by their mass ratio, $q$.  The binaries with $q=1$ appear on the upper strip, positioned $0.752$ magnitudes above their lower strip-single primaries. Isochrones are derived from the 
PARSEC models (see text). 
The red line is the separation between the two strips. The line equation is $-0.44 \times x ^ 2 + 4.3 \times x + 1.37$. It is generated purely based on the $M_{G,0}$ histograms, as explained in section \ref{sec:separating 2 stripes}. For a similar simulation, see Fig.~$1$ in \cite{wallace24}.}
    \label{fig:data with isochrone age 4.5Gyr metallicity 0}
\end{figure}

To show that this is indeed the case, we simulate the CMD location of binary systems at solar metallicity and age of $4.5$ Gyr with different mass ratios, using the PARSEC\footnote{CMD 3.7: http://stev.oapd.inaf.it/cgi-bin/cmd} \citep[e.g.,][]{Bressan12, Chen14, Tang14, Marigo17} tracks
(see Fig.~$1$ of \citet{wallace24}).
Figure~\ref{fig:data with isochrone age 4.5Gyr metallicity 0} seems to support the binary hypothesis, suggesting the upper strip is composed of unresolved binaries with $q\gtrsim 0.7$. 

\subsection{Separating the Binaries From the Single Stars}
\label{sec:separating 2 stripes}

\begin{figure*}	\includegraphics[width=\textwidth, height=10 cm]{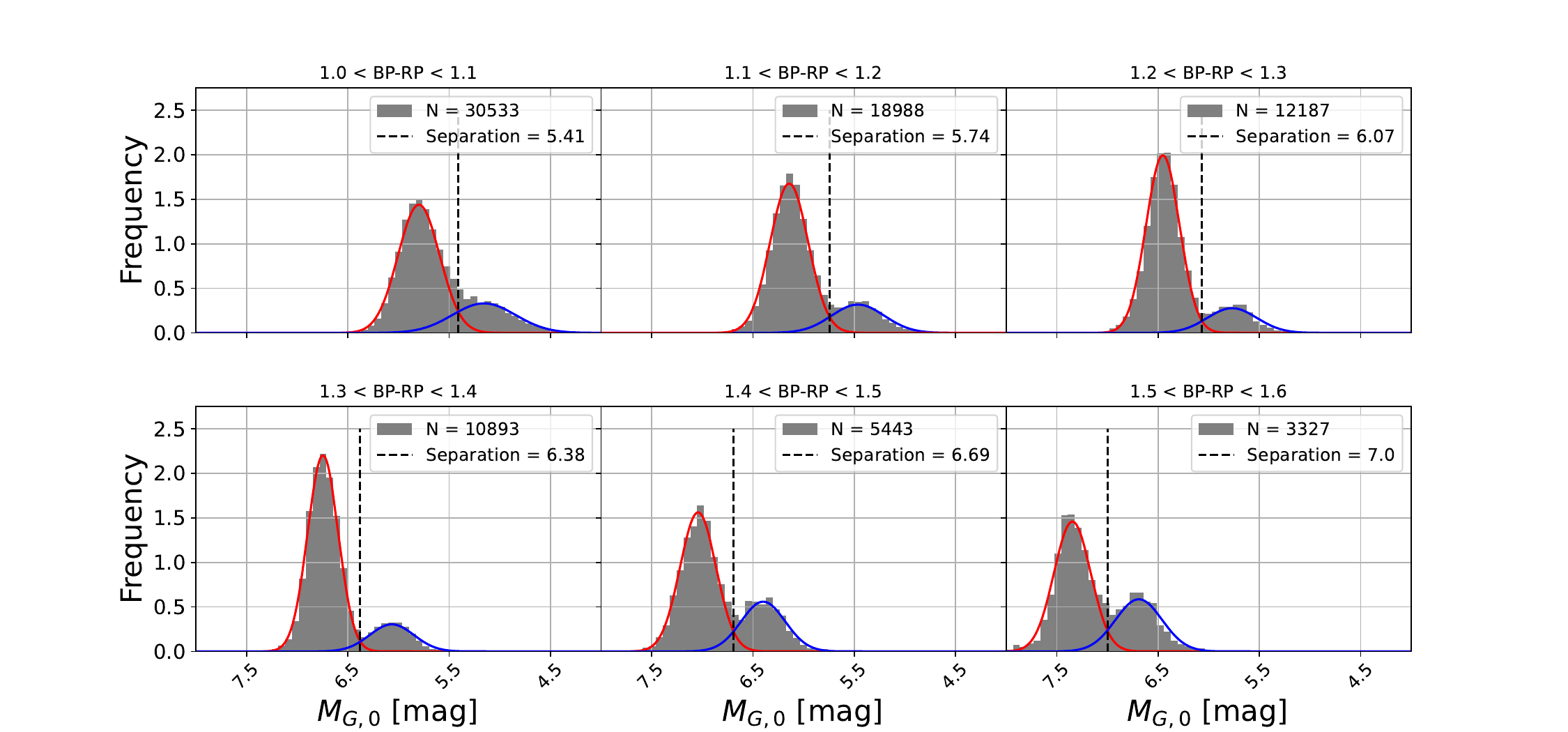}
\caption{Histograms of $M_{G,0}$ per $(G_\mathrm{BP}-G_\mathrm{RP})_0$ bin.}
    \label{fig:BE histograms per bin}
\end{figure*}

To identify the separation between the two populations, we split the $(G_\mathrm{BP}-G_\mathrm{RP})_0$ range into six bins of $0.1$ mag each and plot in Fig.~\ref{fig:BE histograms per bin} the $M_{G,0}$ histogram in each bin. Each of the six histograms resembles a mixture of two Gaussians, corresponding to the lower and upper strips. 

For each histogram, we fit a two Gaussians mixture\footnote{https://scikit-learn.org/stable/modules/generated/sklearn.mixture\\.GaussianMixture.html} and define the separation between the two populations as the intersection of the two Gaussians. Table~\ref{tab:gaussian coeficcients}
lists for each $(G_\mathrm{BP}-G_\mathrm{RP})_0$ bin the parameters of the two Gaussians and the $M_{G,0}$ separation. Note that this splitting between single and binary stars is based only on their CMD location, ignoring their \texttt{vbroad} values. \\
Positioning the six points of separation in the middle of their respective bins, we fit a $2^{nd}$-degree polynomial that separates the two strips, as shown in Fig.~\ref{fig:data with isochrone age 4.5Gyr metallicity 0}.

\begin{table*}
\caption{
Coefficients of Gaussian Mixture 
\label{tab:gaussian coeficcients}}
\begin{center}
\begin{tabular}{lcccccc}
\hline
& Bin 1 & Bin 2 & Bin 3 & Bin 4 & Bin 5 & Bin 6\\
\hline
Separation [mag] & 5.41 & 5.74 & 6.07 & 6.38 & 6.69 & 7.0 \\ 
1st Gaussian Mean [mag] & 5.8 & 6.14 & 6.45 & 6.75 & 7.04 & 7.35  \\ 
1st Gaussian std [mag] & 0.04 & 0.04 & 0.03 & 0.02 & 0.03 & 0.03 \\
2nd Gaussian mean [mag] & 5.16 & 5.46 & 5.77 & 6.07 & 6.4 & 6.69 \\
2nd Gaussian std [mag] & 0.1 & 0.07 & 0.06 & 0.05 & 0.05 & 0.05 \\
Difference between the Means [mag] & 0.64 & 0.68 & 0.68 & 0.68 & 0.64 & 0.66 \\

\hline
\end{tabular}
\end{center}
\end{table*}



\begin{figure*}
    \centering
    \includegraphics[width=\textwidth, height=10 cm]{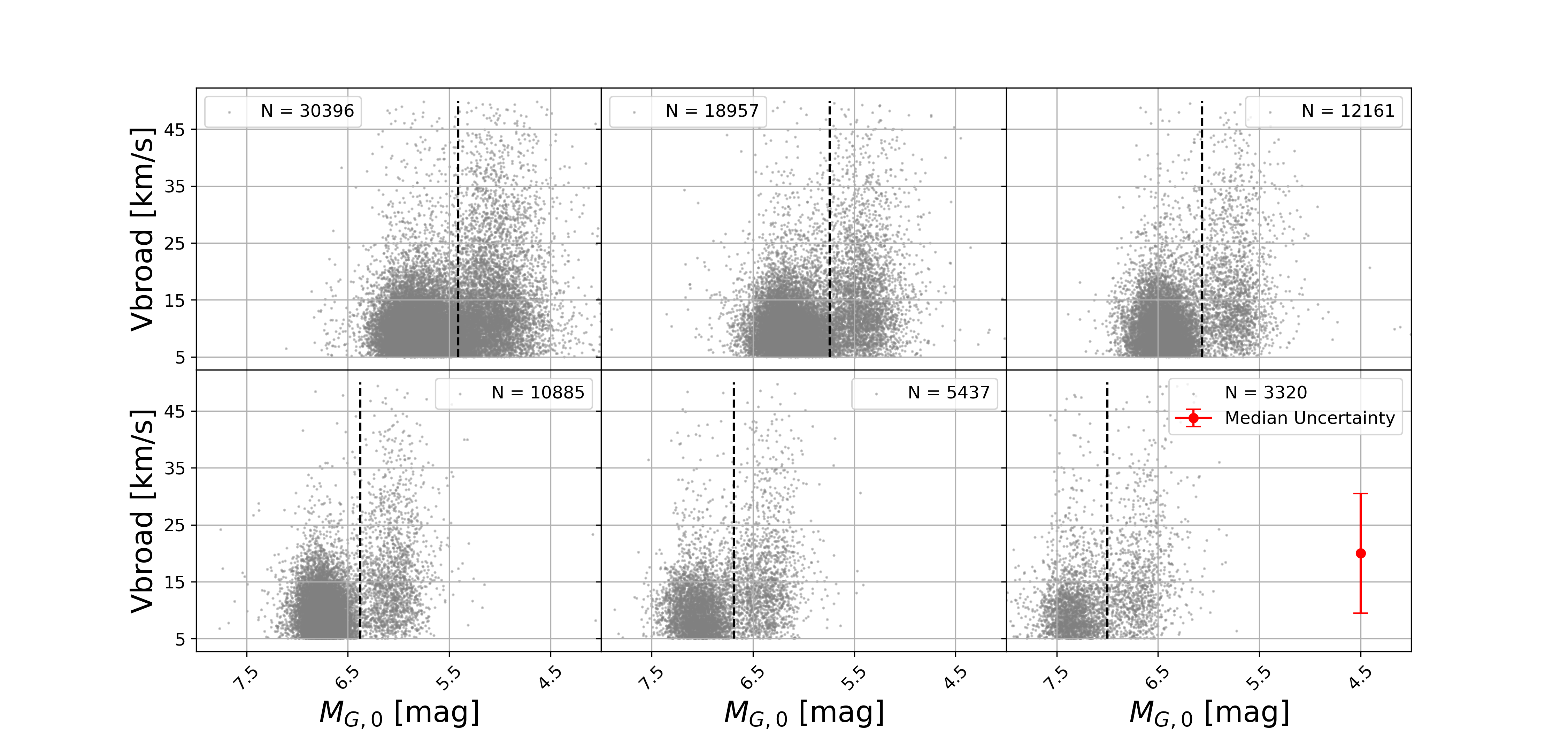}
    \caption{\texttt{Vbroad} vs. $M_{G,0}$. For visual clarity, only $\texttt{vbroad} < 50$ km/s sources are shown. The bottom right panel shows an artificial point with the median uncertainty of the sample.}
    \label{fig:vbroad vs BE}
\end{figure*}

\begin{figure}
    \centering
    \includegraphics[width=\columnwidth]{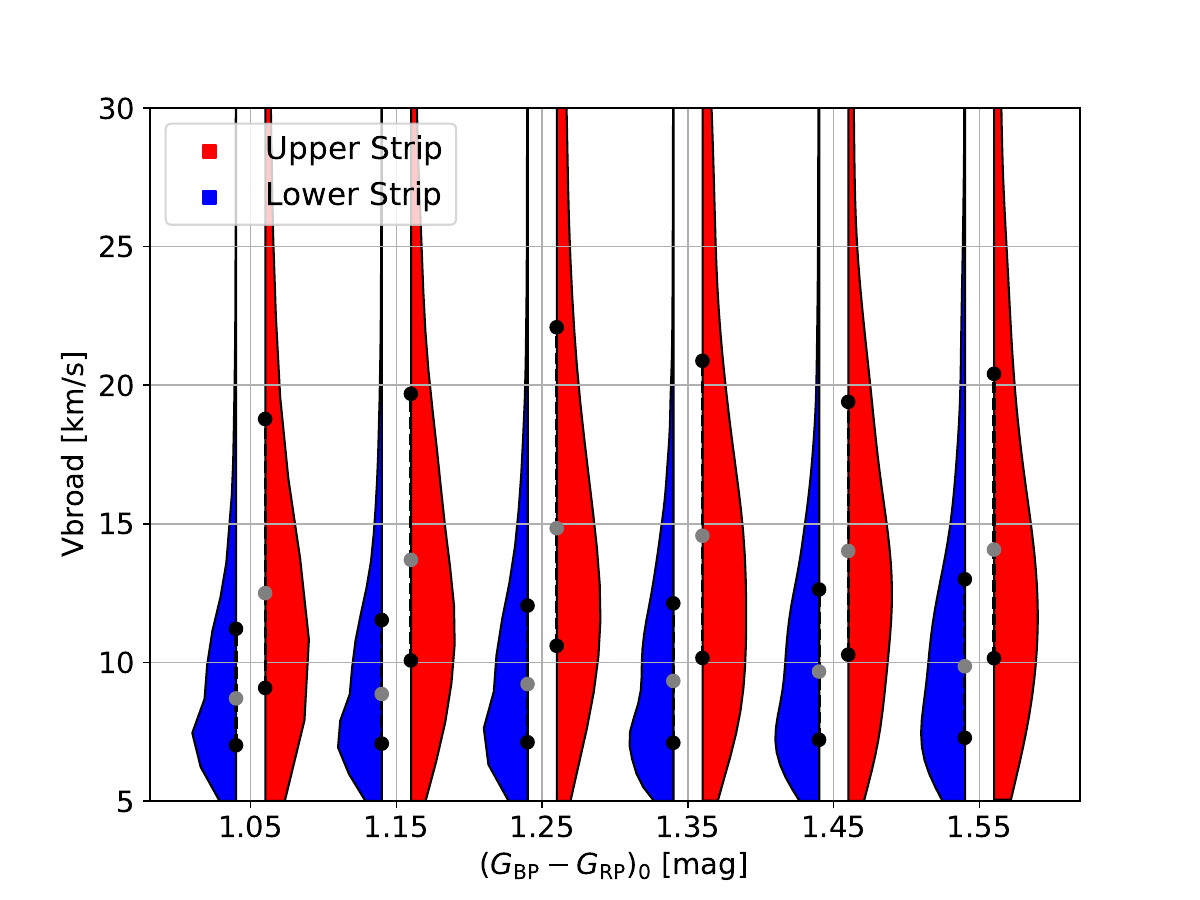}
    \caption{Violin plot for the \gaia \texttt{vbroad} values, divided by colour bin and upper/lower strip. \texttt{Vbroad} is capped here at $30$ km/s for visual clarity; the upper-strip distribution is elongated and extended toward higher \texttt{vbroad} values than the lower ones. Two black dots on each violin represent the $25th$ and $75th$ percentiles; the gray dots mark the median.}
    \label{fig:vbroad violin plot}
\end{figure}

The \texttt{vbroad} values of the two populations are displayed in Fig.~\ref{fig:vbroad vs BE} as a function of the $M_{G,0}$ for the six colour bins. The figure distinctively shows a more extended \texttt{vbroad}  distribution for the binaries. The different distributions are presented as violin plots in Fig.~\ref{fig:vbroad violin plot}. 


\subsection{Comparison with \gaia DR3 MultipleStar Classifier}
\label{sec:logposterior_msc}
It is interesting to compare our binary identification against other indications of binarity.
One such indicator is the multiple star classifier (MSC) of \gaia that infers stellar parameters for bright sources from the low-resolution BP/RP spectra and parallaxes, under the assumption that each of these sources is an unresolved coeval binary system.
Their analysis resulted in a \texttt{logposterior\_msc} parameter\footnote{https://gea.esac.esa.int/archive/documentation/GDR3/Gaia\_archive\\/chap\_datamodel/sec\_dm\_astrophysical\_parameter\_tables/ssec\_dm\\\_astrophysical\_parameters.html\#astrophysical\_parameters-logposterior\_msc} that signifies how likely a system is a binary.

Table \ref{tab:logposterior comparison} compares the \texttt{logposterior} quartiles of the two strips, showing remarkably higher values, with higher probabilities of binarity, for the upper strip, supporting our conjecture.

\begin{table*}
\caption{
Comparison of the \texttt{logposterior} binary indicator for the lower and upper stripe 
\label{tab:logposterior comparison}}
\begin{center}
\begin{tabular}{lcccccc}
\hline
& Lower Strip & Upper Strip\\
\hline
25$^{th}$ percentile & -7573.17 & 430.42\\ 
50$^{th}$ percentile & -4032.42 & 633.50\\ 
75$^{th}$ percentile & -1513.59 & 729.08\\
\hline
\end{tabular}
\end{center}
\end{table*}

\clearpage

\section{GALAH \texttt{Vbroad}}
\label{sec:GALAH}
In this section, we apply our analysis to a different, more accurate, dataset of GALAH \cite{Buder21}, whose DR3 contains $588\,571$ sources; out of which, $588\,464$ have \gaia \texttt{source\_id}. 
After matching based on \texttt{source\_id}, we apply the same filters as described in Section \ref{sec:Identifying 2 Stripes}: $\varpi > 0$ and $\sigma_{\varpi}/\varpi < 10\%$ brings us down to $422\,785$ stars. Now zooming into our two-strips area, $(G_\mathrm{BP}-G_\mathrm{RP})_0$ $\in [1, 1.6]$ and $M_{G,0}$ $\in [4,8]$, leaves us with $32\,675$ sources. Out of those, $32\,430$ have GALAH's \texttt{vbroad} ($4\,529$ of them also have \gaia \texttt{vbroad}).

\begin{figure}
    \centering
    \includegraphics[width=\columnwidth]{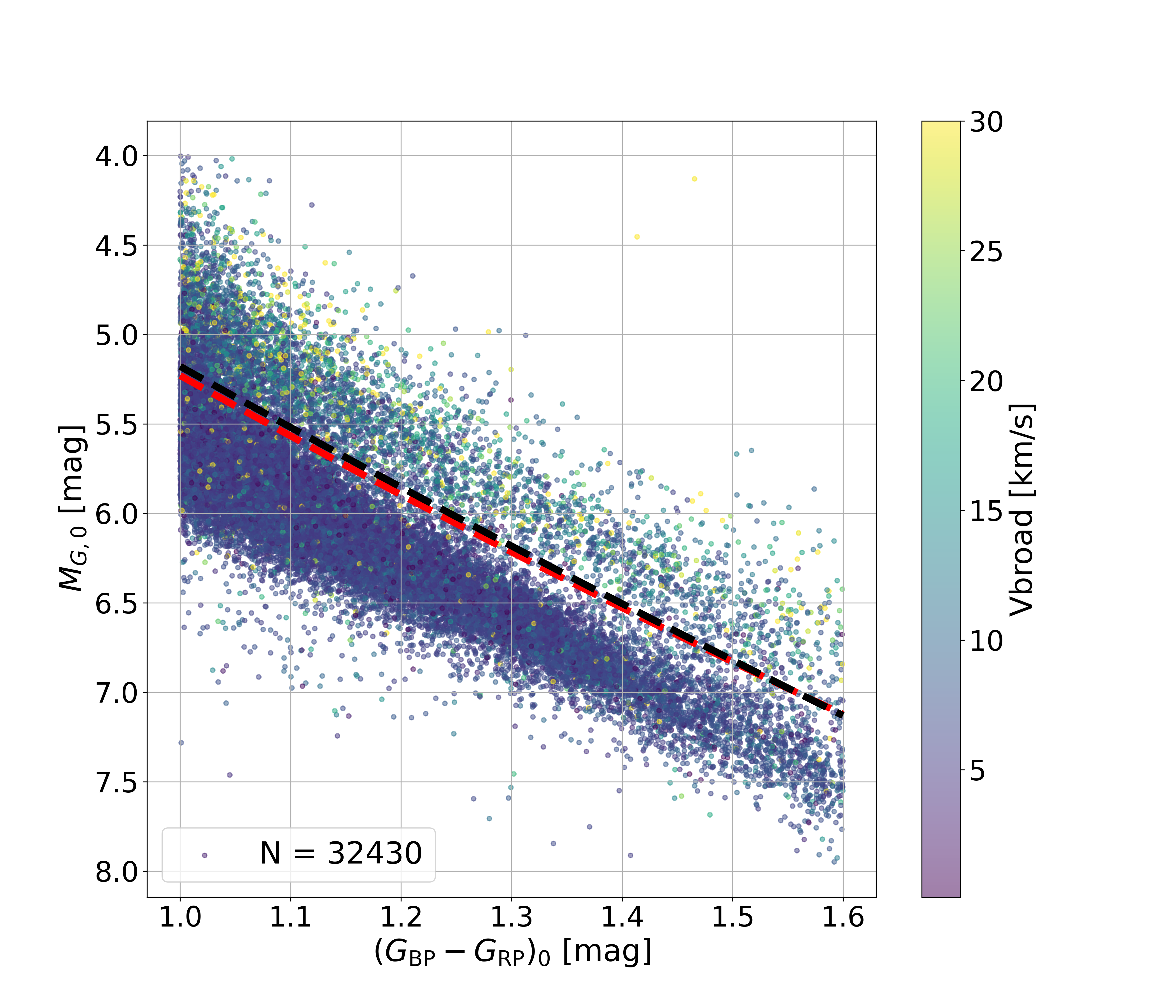}
    \caption{Same as Fig.~\ref{fig:2 stripes color coded} for GALAH \texttt{vbroad} population. The equation for the black line is 
    $y = -0.314 \times x^2 + 4.070 \times x + 1.422$. The red line is the separation derived for the original population, over-plotted here for comparison. The $411$ sources with \texttt{vbroad} above $30$ km/s were rounded to that value; of these, $242$ are on the upper strip.}
    \label{fig:GALAH 2 stripes with 2 lines}
\end{figure}

\begin{figure}
    \centering
    \includegraphics[width=\columnwidth]{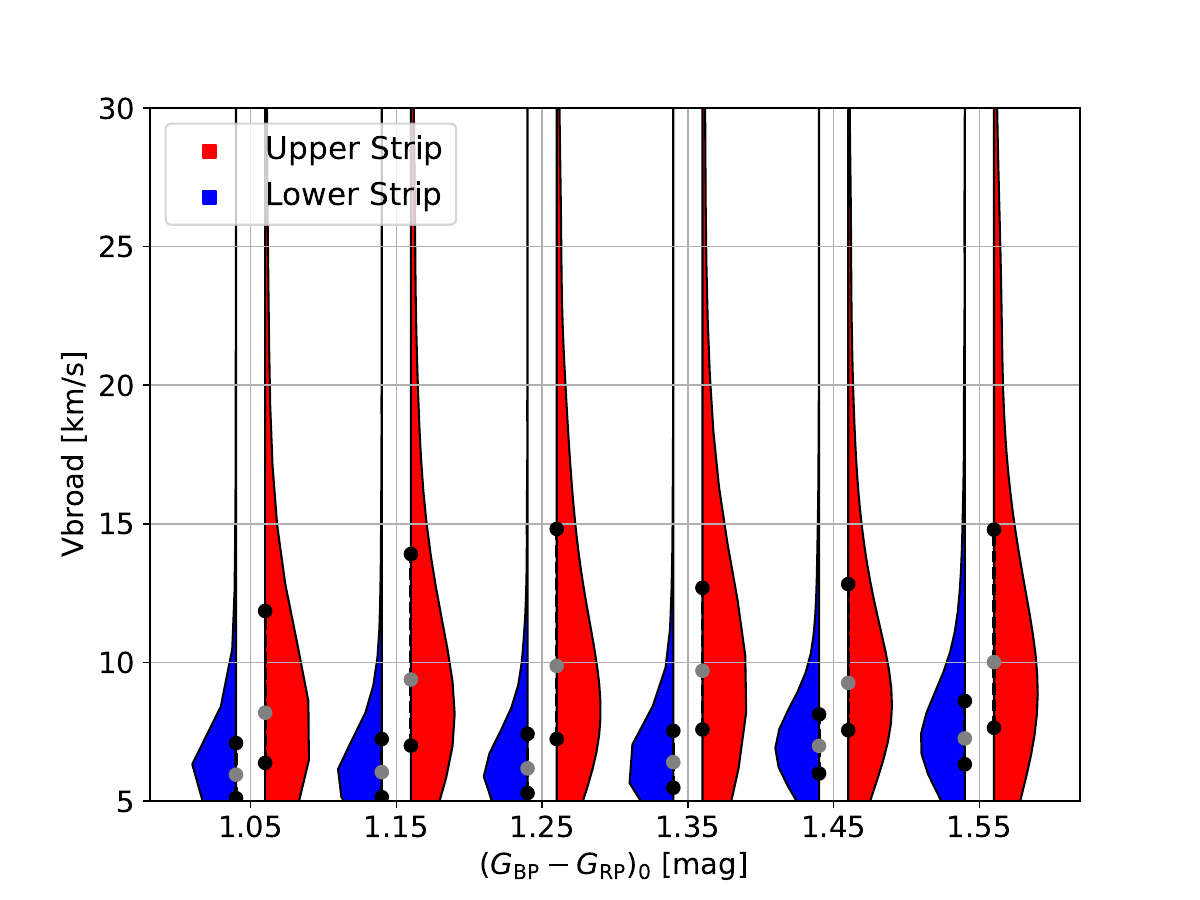}
    \caption{Same as Fig.~\ref{fig:vbroad violin plot}, for GALAH's \texttt{vbroad}. \texttt{Vbroad} is limited at $5$--$30$ km / s for visual clarity. A similar trend is shown between the upper and lower strips. Two black dots are marked on each violin; they represent the $25th$ and $75th$ percentiles; the grey dot marks the median.}
    \label{fig:GALAH vbroad histograms}
\end{figure}

We choose to show only the main figures from the GALAH analysis. Figure~\ref{fig:GALAH 2 stripes with 2 lines} shows the two strips, over-plotted with GALAH separating line, together with the original \gaia line; the two are very similar. Figure~\ref{fig:GALAH vbroad histograms} presents the violin plots, again showing close similarity to Figure~\ref{fig:vbroad violin plot}. The \texttt{logPosterior\_MSC} analysis gave similar results.

\subsection{Example of GALAH spectra with different \texttt{vbroad}}

\begin{figure*}
    \centering
    \includegraphics[width=\textwidth]{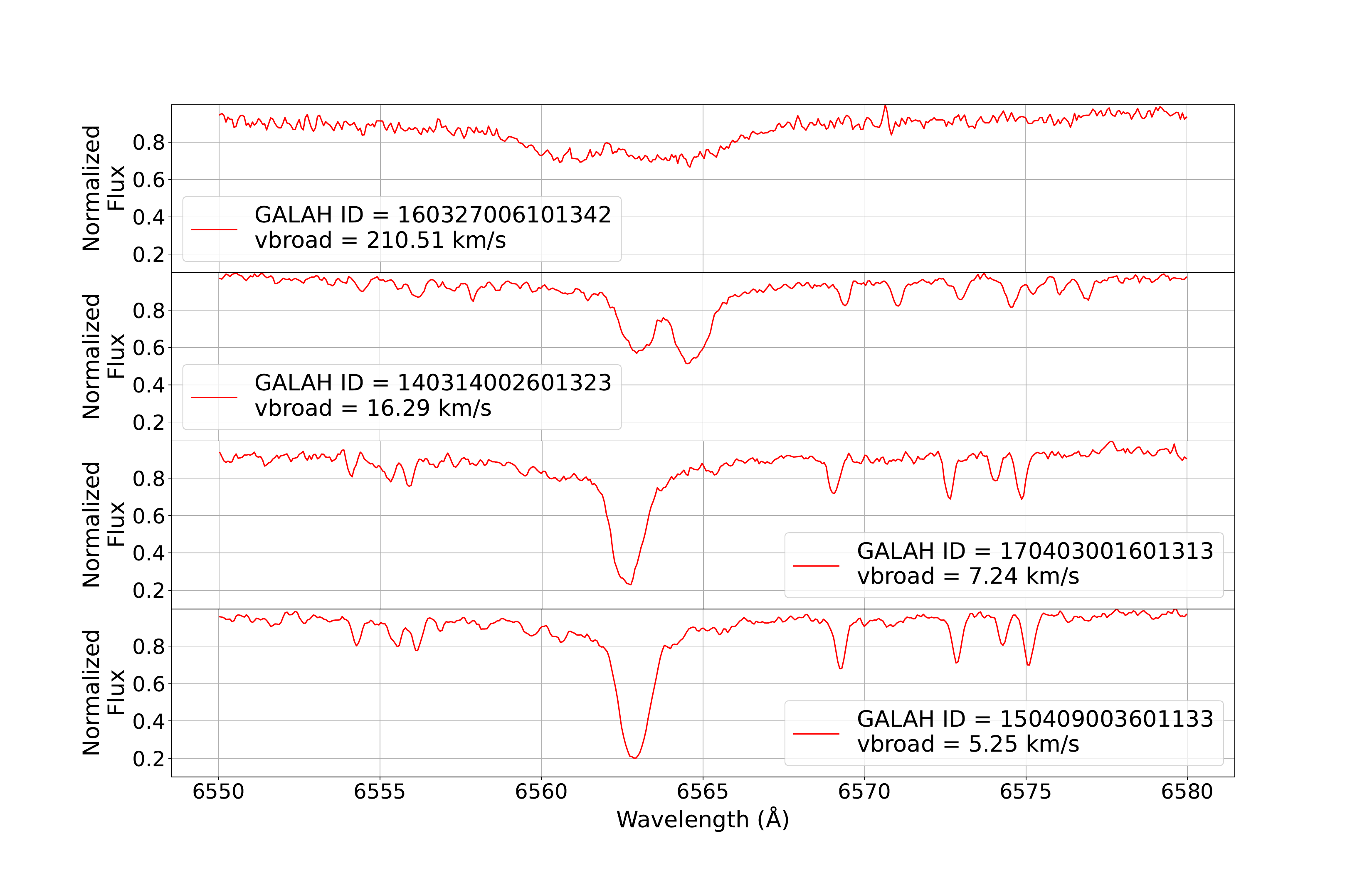}
    \caption{GALAH spectra for four sources, two from the upper strip ($M_{G,0} = 4.74$ for both) and two from the lower one, with a brightness of $M_{G,0} $ of $5.49$ and $5.50$, respectively. The colour of the sources is $(G_\mathrm{BP}-G_\mathrm{RP})_0 = 1.00$--$1.01$. The brighter sources show much broader absorption lines.} 
    \label{fig:GALAH spectra}
\end{figure*}

Figure \ref{fig:GALAH spectra} shows an example of four GALAH spectra of the H$\alpha$-line region for sources with similar colour temperature 
($5\,080$--$5\,220~\mathrm{K}$, taken from \texttt{teff\_gspphot}\footnote{https://gea.esac.esa.int/archive/documentation/GDR3/\\Gaia\_archive/chap\_datamodel/sec\_dm\_main\_source\_catalogue\\/ssec\_dm\_gaia\_source.html\#gaia\_source-}). Two of the sources, which are twice as bright as the other two, are suspected binaries, with \texttt{vbroad} values that  are much larger than those of the two single stars. One can see that the  H$\alpha$-line is split in those two spectra, for the ID=140314002601323 source in particular, consistent with the binary conjecture.  


\section{Discussion}
\label{sec:discussion}
This study considers \NvbroadTwoStripes~\gaia late-type MS sources with derived \texttt{vbroad} values, the CMD of which displays a two-strip structure, as seen already by \cite{freund24}, \cite{phillips24} and  \cite{wallace24}. The objects in the upper strip, identified by their $M_{G,0}$ excess, show a higher \texttt{vbroad}-value distribution. 
Very similar behavior has been seen in the GALAH data. 

We have shown that many of the upper strip stars are composed of unresolved binaries by simulating isochrones of binaries with different mass ratios. The simulations show that, regardless of the color, the upper strip includes binaries with mass ratios larger than about $0.7$.

The \texttt{vbroad} derivation is based on the \gaia RVS relatively low-resolution spectra, and therefore the \texttt{vbroad} values are usually associated with large uncertainties \citep[][]{Fremat23}. Nevertheless, we were able to identify the binary effect on these values with high significance, because of the large number of sources available for the analysis.

We suggest that the larger \texttt{vbroad} values of many of the unresolved binaries originate from the presence of the lines of the secondary star in the spectra of those binaries. The RV's of the two components of the binary systems are separated by their orbital motion, but cannot be resolved by the RVS spectra. Therefore, the \texttt{vbroad}
pipeline measures a larger line width. Obviously, the RV difference between the two components depends on the orbital period and the binary inclination and phase at the time of each measurement, parameters that are not available. Nevertheless, the additional component induces an extra width to the line profile and consequently to the \texttt{vbroad} estimation of the unresolved binaries. 

The \texttt{vbroad} excess of the unresolved binaries is on the order of $10$ km/s. For a binary with a total mass of $1 M_{\odot}$, a relative velocity of $\sim 10$ km/s indicates an orbital period of $\sim 30$ years and $\sim 10$ au, assuming an orbital inclination close to $90^{\circ}$. Such a binary was impossible to detect as an astrometric binary in DR3 \citep[e.g.,][]{arenou23}, on the one hand, and cannot be resolved by \gaia \citep[e.g.,][]{catro-ginard24} on the other hand. \cite{arenou23} classified $12\%$ of the upper strip sources while only $4\%$ of the lower strip sources as binaries, strengthening the suggestion about the high frequency of binarity in the upper strip.  

We note that $\sim 20\%$ of the \gaia and GALAH \texttt{vbroad} sources are on the upper strip. Assuming a high frequency of binaries in these upper strips is consistent with the estimation that about $50\%$ of $0.1$--$1 M_{\odot}$ stars are members of multiple-star systems \citep[e.g.,][]{Raghavan10,duchene13,moe17}, taking into account that only unresolved high mass-ratio binaries reside on the upper strip. 

In the future, one will be able to detect an increasing number of these binaries as \gaia astrometric binaries, as the observational time span of the space mission increases. Further, with the TODCOR technique \citep{todcor94}, one could try and identify the secondary spectrum in the RVS spectrum, like was done for the GALAH project \cite[][]{traven20}, or in the Gaia BP/RP spectra \citep{XP23}. These analyses will bring new opportunities for a better understanding of the true characteristics of the binary population of late-type stars. 

\begin{acknowledgements}
We are indebted to the referee for thoughtful advice
that significantly improved the previous version of the paper. This work has made use of data from the European Space Agency (ESA) mission
{\it Gaia} (\url{https://www.cosmos.esa.int/gaia}), processed by the {\it Gaia}
Data Processing and Analysis Consortium (DPAC,
\url{https://www.cosmos.esa.int/web/gaia/dpac/consortium}). Funding for the DPAC
has been provided by national institutions, in particular the institutions
participating in the {\it Gaia} Multilateral Agreement.
This work also made use of the Third Data Release of the GALAH Survey (Buder et al. 2021). The GALAH Survey is based on data acquired through the Australian Astronomical Observatory, under programs: A/2013B/13 (The GALAH pilot survey); A/2014A/25, A/2015A/19, A2017A/18 (The GALAH survey phase 1); A2018A/18 (Open clusters with HERMES); A2019A/1 (Hierarchical star formation in Ori OB1); A2019A/15 (The GALAH survey phase 2); A/2015B/19, A/2016A/22, A/2016B/10, A/2017B/16, A/2018B/15 (The HERMES-TESS program); and A/2015A/3, A/2015B/1, A/2015B/19, A/2016A/22, A/2016B/12, A/2017A/14 (The HERMES K2-follow-up program). We acknowledge the traditional owners of the land on which the AAT stands, the Gamilaraay people, and pay our respects to elders past and present. This paper includes data that has been provided by AAO Data Central (datacentral.org.au).
\\
We are indebted to the \gaia CU6 team that released the \texttt{vbroad} values for such a large sample, enabling this study.
\end{acknowledgements}

\bibliographystyle{aa}
\bibliography{main} 

\end{document}